\documentclass[12pt]{iopart}%
\usepackage{color}
\usepackage{graphicx}
\usepackage{iopams}

\def\ad{a^{\dagger}}

\begin{document}
\title[Preparing and Probing Atomic Majorana Fermions in Optical Lattices]{Preparing and Probing Atomic Majorana Fermions and Topological Order in Optical Lattices}
\author{C. V. Kraus$^{1,2}$, S. Diehl$^{1,2}$, P. Zoller$^{1,2}$, M.A. Baranov$^{1,2}$}

\address{$^{1}$Institute for Quantum Optics and Quantum Information of the Austrian Academy of Sciences, A-6020 Innsbruck, Austria}
\address{$^{2}$Institute for Theoretical Physics, University of Innsbruck, A-6020 Innsbruck, Austria}
\address{$^{3}$NRC ``Kurchatov Institute'', Kurchatov Square 1, 123182 Moscow, Russia}

\begin{abstract}
We introduce a one-dimensional system of fermionic atoms in an optical lattice
whose phase diagram includes topological states of different symmetry classes
with a simple possibility to switch between them. The states and topological
phase transitions between them can be identified by looking at their
zero-energy edge modes which are Majorana fermions. We propose several
universal methods of detecting the Majorana edge states, based on their
genuine features: zero-energy, localized character of the wave functions, and
induced non-local fermionic correlations.

\end{abstract}
\maketitle

\section{Introduction}
There is at present growing interest in topological phases of matter --
fermionic insulators and superconductors with a gapped excitation spectrum and
nonlocal topological order (TO). TO is characterized by a topological
invariant taking discrete values \cite{Schnyder_Ludwig,
Schnyder_Ludwig_PRB,Kitaev_AIP, Qi, Zirnbauer, Altland_Zirnbauer}. An immediate
consequence of TO is the existence of robust zero-energy modes localized at
defects, edges, and interfaces between different topological phases
\cite{Teo_Kane, Gurarie,Bulk_edge_Gurarie}. The number of these states and
their properties are directly linked to TO and, therefore, can be viewed as a
probe of the topological state providing access to the corresponding phase
diagram. A striking example are zero-energy Majorana fermions, as discussed in
the context of topological quantum computing \cite{Sau_TQC,Alicea_TQC,
Nayak_TQC}.

Most of the developments and proposals to realize and study topological phases
in general, and Majorana fermions in particular, have been related to
condensed matter systems \cite{Probing_KaneFu,Probing_Fu, Probing_Sau,
Beenakker, Stanescou_DasSarma}, with a first observation of Majorana edge
modes in a hybrid superconductor-semiconductor nanowire device recently
reported in Ref.~\cite{MajoranaExperiment}. On the other hand, there are
several promising proposals to realize topological phases with quantum
degenerate gases of atoms and molecules~\cite{Goldman_Lewenstein,
Shlayapnikov_Cooper, DasSarma_opticallattice,DasSarma_flatband, Probing_Liang,
Diehl_Natphys}, with questions of detection of such phases presently in the
focus of interest~\cite{Probing_Cooper_Auerbach,Spielman_Chern, Ripoll_Pachos,
Spielman_Viewpoint}. Below we will outline a measurement scenario for
topological phases and phase transitions by \emph{monitoring signatures of
Majorana fermion edge-states} in atomic systems. Our analysis builds directly
on recent experimental advances such as single site addressing and
measurements in optical lattices \cite{Bloch_singlespin, Greiner_singlesite}
in combination with traditional time of flight and spectroscopic techniques,
and complements the recent proposals to detect topology in the bulk as
proposed in \cite{Spielman_Chern, Ripoll_Pachos,Spielman_Viewpoint}.

To illustrate our ideas for detection we introduce a simple, but
experimentally realistic example of a zig-zag chain (c.f.~Fig.~1a). This model
is an extension of the familiar Kitaev model of spin-less fermions coupled to
a BCS-reservoir with the additional feature of next-to-nearest neighbor
couplings (see below), and can be realized with cold atoms generalizing ideas
outlined in Ref.~\cite{Probing_Liang}. The model has a remarkably rich
phase diagram, allowing different topological states in two symmetry classes
supporting Majorana edge modes, and provides an ideal playground to
demonstrate the presence of TO and the topological phase transitions via
related Majorana edge states. We propose a fully reproducible preparation of
an initial state for Majorana fermions\ and detection with atomic measurement
techniques, including (i) zero-energy, (ii) localization near the edge and
(iii) induced non-local fermionic correlations, which, together, allow for an unambiguous probe of Majorana fermions. While the present example is
1D, the techniques described below have a straightforward extension to 2D, as
illustrated by discussion of a $p_{x}+ip_{y}$ superfluid.
\begin{figure}[ptb]
\includegraphics[width=0.95\columnwidth]{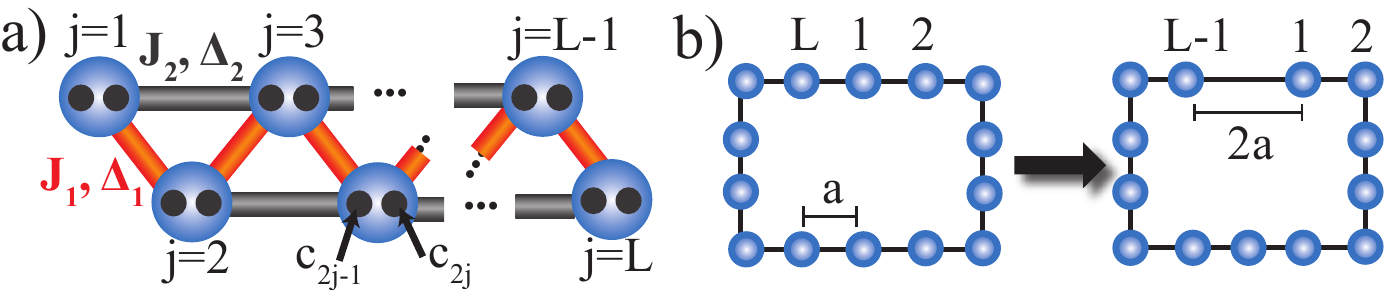} \caption{
a) A zig-zag chain of spinless fermionic atoms according to the Hamiltonian~$H$
with NN (red) and NNN couplings (grey). $c_{2j-1}$ and $c_{2j}$ denote
Majorana operators on lattice sites $j=1$ to $L$ (see text). b) Preparation of
the initial state of Majorana fermions by adiabatically cutting the closed
chain.}%
\label{fig:Zigzag_main}%
\end{figure}

\section{The Zig-zag Chain}
In the following Section we discuss the Zig-zag Chain, a one-dimensional ($1D$) model system of spinless fermions which allows for a variety of topological phases, thus providing an ideal playground for testing our detection techniques. We start in Subsec.~\ref{ref:Sec_model} introducing the model and presenting its topological phase diagram. The reader interested in the derivation of these results is referred to Subsec.~\ref{Sec_phasediagram} . Further, we review the ideas of ~\cite{Probing_Liang} in Subsec.~\ref{Sec_pairingterm}, explaining how the zig-zag chain can be realized in a cold atom implementation. 

\subsection{The Model System and its Properties}\label{ref:Sec_model}
We consider single-component fermions on a finite $1D$ chain of size $L$ with Hamiltonian $H=H_{1}%
+H_{2}+H_{\mu}$ (c.f.~Fig.~1a). Here, $H_{\mu}=-\mu\sum_{j=1}^{L}%
a_{j}^{\dagger}a_{j}$, and
\begin{equation}\label{eq:H_zigzag}
H_{\alpha}=\sum_{j=1}^{L-\alpha}\left[  -J_{\alpha}a_{j}^{\dagger}a_{j+\alpha
}+\Delta_{\alpha}a_{j}a_{j+\alpha}+h.c.\right] .
\end{equation}
where $a_{i}$ and $a_{i}^{\dagger}$ are fermionic operators ($i=1,\ldots,L$),
$J_{\alpha}\geq0$ and $\Delta_{\alpha}=\left\vert \Delta_{\alpha}\right\vert
e^{i\phi_{\alpha}}$ are nearest-neigbour (NN, $\alpha=1$) and
next-to-nearest-neighbour (NNN, $\alpha=2$) hopping and pairing amplitudes,
respectively, and $\mu$ is the chemical potential. As shown in ~\cite{Probing_Liang} and reviewed in Subsec.~\ref{Sec_pairingterm}, the pairing terms in
$H_{\alpha}$ are obtained by a Raman induced dissociation of Cooper pairs (or
Feshbach molecules) forming an atomic BCS reservoir, with
$\mu$ a Raman detuning. In Fig.~\ref{fig:Zigzag_main}a we represent the chain
as a zig-zag, which leads naturally to a cold atom implementation with optical
lattices allowing control of the relative strength of NN and NNN amplitudes
by changing the zig-zag geometry.

Although $H$ can be viewed as a Hamiltonian of two coupled Kitaev
chains~\cite{Kitaevchain} (with odd or even sites), the resulting topological
phase diagram is substantially richer. Since the coupling parameters can be
changed in a real-time experiment, this setup gives rise to a platform for
exploring topological properties of matter, including topological phase
transitions and Majorana fermions. We summarize here only the main features of the topological phase diagram, and refer the interested reader to Subsec.~\ref{Sec_phasediagram}

The topological symmetry class \cite{Schnyder_Ludwig} of $H$ depends crucially on the relative phase
$\phi=\phi_{1}-\phi_{2}$ of the pairing amplitudes (one of the phases, say,
$\phi_{2}$,$\,$can be gauged away by redefining the operators $a_{j}$). For
$\phi=0,\pi$ $(\mathrm{mod}\,2\pi)$ it is the class
\textrm{BDI} (Cartan) with time-reversal, particle-hole, and chiral symmetries, for all
other values of $\phi$ it is class \textrm{D} with only particle-hole symmetry. Both classes allow topologically nontrivial states characterized
by the integer-valued winding number $\nu\in\mathbb{Z}$ in the class
\textrm{BDI} and by the Chern parity number $P_{C}=0,1\in\mathbb{Z}_{2}$ in
the class \textrm{D}.
\begin{figure}[ptb]
\includegraphics[width =0.95\columnwidth]{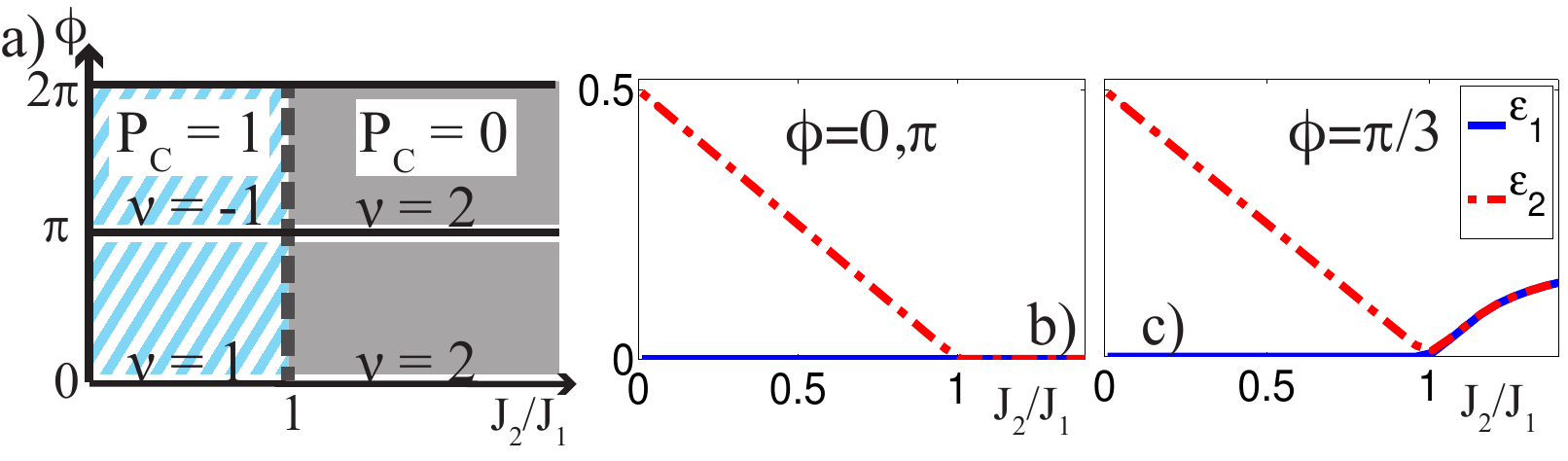}
\caption{a) Phase diagram corresponding to the Hamiltonian~$H$
for $\mu=0$, $|\Delta_{\alpha}|=J_{\alpha}e^{i\phi}$. b) The two lowest doubly
degenerate eigenvalues $\varepsilon_{1,2}$ of $H$ as a function of the ratio
$J_{2}/J_{1}$: If $\phi=0,\pi$ $(\mathrm{mod}\,2\pi)$ (\textrm{BDI} class)
there are one or two such modes (b), for all other values of $\phi$
(\textrm{D} class), there is at most one zero mode (c) (see also Subsec.~\ref{Sec_phasediagram}).}%
\label{fig:Phase_diagram_allphi}%
\end{figure}
The topological phase diagram contains states with $\nu=0,\pm
1,\pm2$ for the \textrm{BDI} class, and with $P_{C}=0,1$ for the \textrm{D}
class. Fig.~\ref{fig:Phase_diagram_allphi}a summarizes results for the
simplest case $\mu=0$ and $\left\vert \Delta_{\alpha}\right\vert =J_{\alpha}$,
where the phase diagram is characterized by the relative phase $\phi$ and the
ratio $J_{2}/J_{1}$. We find that $\nu  =2, P_{C}=0$ for $J_{2}>J_{1}$, else one
has $P_{C}=1$ and $\nu=+1$ or $-1$ for $\phi=0$ and $\pi$, respectively.
Therefore, for both symmetry classes, the point $J_{1}=J_{2}$ corresponds to a
topological phase transition, where the bulk excitations become \emph{gapless} (see Subsec.~\ref{Sec_phasediagram}).

The presence of a nontrivial TO in the bulk leads to zero-energy modes, which
in our case are Majorana fermions, at the interfaces between states with
different TO. The number of these states is related to the difference of the
corresponding topological invariants \cite{Gurarie, Bulk_edge_Gurarie}. Using
the hermitian Majorana representation, where $c_{2j-1}=a_{j}^{\dagger}+a_{j}$
and $c_{2j}=(-i)(a_{j}^{\dagger}-a_{j})$ obey $\{c_{j},c_{l}\}=2\delta_{jl}$,
the quadratic Hamiltonian reads $H=i\sum_{j,l}A_{jl}c_{j}c_{l}$ with real
$A=-A^{T}$. Diagonalizing $A$ for a finite chain we identify the zero-energy
modes and associated Majorana edge-mode operator $\gamma_{E}=\sum_{k}%
v_{Ek}c_{k}$ with (real) coefficients $v_{Ek}$ localized at the interface with
some localization length $l_{\mathrm{loc}}$. [To be precise, the energy of
these modes $\sim\exp(-L/l_{\mathrm{loc}})$ approaches zero for $L\rightarrow
\infty$.] Thus, by detecting Majorana zero-energy edge modes and their number
we can access the full topological phase diagram of our model:

For the \textrm{BDI} class we have in total two such modes for $J_{1}>J_{2}$:
$\gamma_{L}=c_{1}$ and $\gamma_{R}=c_{2L}$ for the left and the right edge,
respectively, and four modes for $J_{1}<J_{2}$. The two additional zero modes
are exponentially decaying inside the bulk (see
Fig.~\ref{fig:Phase_diagram_allphi}b and Subsec.~\ref{Sec_phasediagram}). For the \textrm{D}
class there are two zero-energy modes for $J_{1}>J_{2}$, which decay
exponentially inside the chain, and no such modes for $J_{1}<J_{2}$
(Fig.~\ref{fig:Phase_diagram_allphi}c).

The presence of Majorana zero-energy modes comes along with a degeneracy of
the ground state: The dimension of the ground-state subspace is $2^{N_{M}/2}$,
where $N_{M}$ is the number of the Majorana modes (always even in our case). A
state in this subspace generically has correlations between Majorana operators
from different edges, resulting in non-local fermionic correlations (in
contrast to local correlations in the bulk). For the case with only two
Majorana modes ($\nu=1$ or $P_{C}=1$), the two ground states $\left\vert
G_{\pm}\right\rangle $ have even ($+$) or odd ($-$) number of fermions,
respectively, and the nonlocal correlations $\left\langle G_{\pm}\right\vert
\gamma_{L}\gamma_{R}\left\vert G_{\pm}\right\rangle =\pm i$ are related to the
fermionic parity, in full analogy to Ref. \cite{Kitaevchain}.

\subsection{Derivation of the Topological Phase Diagram}\label{Sec_phasediagram}
In the following we present a detailed derivation of the topological phase diagram presented in Subsec.~\ref{ref:Sec_model}. According to the general tenfold classification scheme~\cite{Schnyder_Ludwig, Kitaev_AIP}, the topological class to which a given Hamiltonian belongs to, is
determined by the invariance properties of the Hamiltonian under
time-reversal, particle-hole (or charge conjugation), and chiral (or
sublattice) symmetry. This class specifies then, whether the states with a
nontrivial topological order could exist and provides the corresponding
topological invariant to distinguish them.

For a translationally invariant $1D$ spinless Bogoliubov-de Gennes lattice Hamiltonian
in the quasimomentum basis%
\begin{equation}
H=\sum_{k\in\mathrm{BZ}}\Psi_{k}^{\dagger}\mathcal{H}_{k}\Psi_{k},
\end{equation}
where%
\begin{equation}
\mathcal{H}_{k}=%
\left(\begin{array}{cc}
\xi_{k} & \Delta_{k}\\
\Delta_{k}^{\ast} & -\xi_{k}%
\end{array}\right)
,\quad\Psi_{k}=%
\left(\begin{array}{c}
a_{k}\\
a_{-k}^{\dagger},%
\end{array}\right)
\end{equation}
with $a_{k}=L^{-1/2}\sum_{k}e^{ikj}a_{j}$, the invariances are equivalent to
the following conditions:%
\[
U_{T}^{\dagger}\mathcal{H}_{k}U_{T}=\mathcal{H}_{-k}^{\ast},%
\]
for the time-reversal,%
\[
U_{C}^{\dagger}\mathcal{H}_{k}U_{C}=-\mathcal{H}_{-k}^{\ast},%
\]
for the particle-hole, and%
\[
\Sigma^{\dagger}\mathcal{H}_{k}\Sigma=-\mathcal{H}_{k},%
\]
for the chiral symmetries, respectively, where $U_{T}$, $U_{C}$, and $\Sigma$
are unitary matrices satisfying $U_{T}U_{T}^{\ast}=\pm1$, $U_{C}U_{C}^{\ast
}=\pm1$, and $\Sigma^{2}=1$.

For the Hamiltonian $H$ of Eq. (\ref{eq:H_zigzag}) one has $\xi_{k}=-(J_{1}\cos
k+J_{2}\cos2k)-\mu/2$ and $\Delta_{k}=-i(\Delta_{1}\sin k+\Delta_{2}\sin2k)$,
such that the matrix $\mathcal{H}_{k}$ is%
\[
\mathcal{H}_{k}=\vec{h}_{k}\cdot\vec{\sigma},%
\]
where $\vec{\sigma}=\{\sigma_{x},\sigma_{y},\sigma_{z}\}$ is the vector of
Pauli matrices and $\vec{h}_{k}^{T}=[\mathrm{Re}, -\mathrm{Im}(\Delta_{k})
(\Delta_{k}),\xi_{k}]$. The excitation energies are $E_{k}=2\left\vert \vec
{h}_{k}\right\vert =2\sqrt{\xi_{k}^{2}+\left\vert \Delta_{k}\right\vert ^{2}}$.

For generic $\Delta_{1}=\left\vert \Delta_{1}\right\vert e^{i\phi_{1}}$ and
$\Delta_{2}=\left\vert \Delta_{2}\right\vert e^{i\phi_{2}}$, only the
particle-hole symmetry condition is fulfilled with $U_{C}=\sigma_{x}$ (as it
should be for a general Bogoliubov-de Gennes Hamiltonian), and, therefore, the
Hamiltonian $H$ belongs to the class \textrm{D}. For $\phi\equiv\phi_{1}%
-\phi_{2}=0$ or $\pi$ $\mathrm{mod} 2\pi)$, that is $\Delta
_{k}=-ie^{i\phi_{2}}(\varepsilon\left\vert \Delta_{1}\right\vert \sin
k+\left\vert \Delta_{2}\right\vert \sin2k)$ with $\varepsilon=\exp(i\phi
)=\pm1$, the conditions for the time-reversal and chiral symmetries are also
satisfied with $U_{T}=\mathrm{diag}(e^{-i\phi_{2}},e^{i\phi_{2}})$ and
$\Sigma=\sigma_{x}U_{T}$, and the Hamiltonian $H$ belongs to the chiral
\textrm{BDI} class. Note that the phase $\phi_{2}$ can be gauged away by
$a_{k}\rightarrow e^{-i\phi_{2}/2}a_{k}$ such that $\mathrm{Re}(\Delta_{k})=0$
and the vector $\vec{h}_{k}$ belongs to the $yz$-plane for all $k$.
Geometrically speaking, for the \textrm{BDI} class the vectors $\vec{h}_{k}$
for all $k$ are in the same plane, which is the $yz$-plane for $\phi_{2}=0$.

\begin{figure}[t]
\begin{center}
\includegraphics[width=0.8\columnwidth]{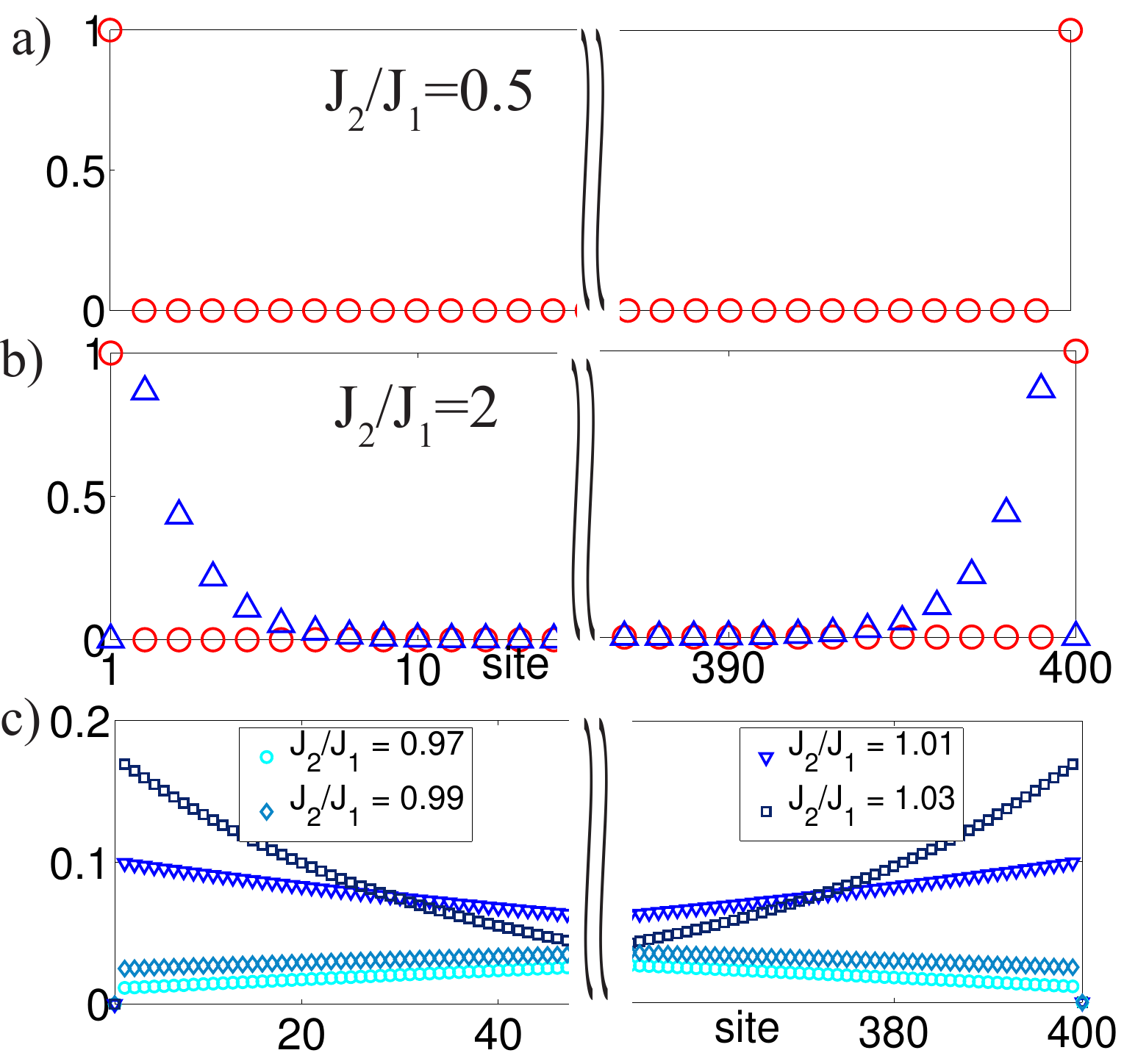}
\caption{Spatial distribution of the zero energy modes of $H$ for $J_{\alpha} = |\Delta_{\alpha}|$ and $\phi = 0$. a) If $J_2<J_1$ we have two zero energy modes that are located at the boundary. b) If $J_2 > J_1$, two additional zero energy modes with an exponential decay appear. c) These modes extend more and more over the lattice when we approach the transition point $J_2 = J_1$.\label{fig:Edgemodes_BDI}}
\end{center}
\end{figure}

For the chiral \textrm{BDI} class in $1D$, different topological states are
classified by the integer valued winding number $\nu\in\mathbb{Z}$ defined in the following way (see, for example, [7]): By gauging away
the phase $\phi_{2}$ and using a unitary transformation $U=\exp(i\pi\sigma
_{y}/4)$, the Hamiltonian can be transformed into a canonical form%
\begin{equation}
\mathcal{H}_{k}=%
\left(\begin{array}{cc}
0 & q_{k}\\
q_{k}^{\ast} & 0
\end{array}\right)
=\left\vert q_{k}\right\vert
\left(\begin{array}{cc}
0 & e^{i\varphi_{q_{k}}}\\
e^{-i\varphi_{q_{k}}} & 0
\end{array}\,\right)
\end{equation}
with $q_{k}=$ $\xi_{k}+\Delta_{k}$ and $\varphi_{q_{k}}$ being the phase of
$q_{k}$. Then
\[
\nu=\frac{1}{2\pi}\int_{-\pi}^{\pi}dk\frac{d\varphi_{q_{k}}}{dk}%
=\varphi_{q_{k=\pi}}-\varphi_{q_{k=-\pi}}\in\mathbb{Z}
\]
Alternatively, the winding number can be defined using the unit vector $\vec
{n}_{k}=\vec{h}_{k}/\left\vert \vec{h}_{k}\right\vert $ in the $yz$-plane
($\phi_{2}$ is gauged away)%
\[
\nu=\frac{1}{2\pi}\int_{-\pi}^{\pi}\hat{e}_{x}\cdot(\vec{n}_{k}\times
\partial_{k}\vec{n}_{k})dk\in\mathbb{Z},
\;
\]
as number of times the vector $\vec{n}_{k}$ wraps around the unit circle when
$k$ goes around the Brioullin zone (here $\hat{e}_{x}$ is the unit vector
along the $x$-axis). Note that as long as $\left\vert \vec{h}_{k}\right\vert
\neq0$ (or, equivalently, $\left\vert q_{k}\right\vert \neq0$), which
corresponds to the gapped spectrum of the Hamiltonian, the winding number is
well-defined and, therefore, can change its value only when the gap closes.
This gap-closing condition is the necessary one for the topological phase transition.

\begin{figure}[t]
\begin{center}
\includegraphics[width=0.5\columnwidth]{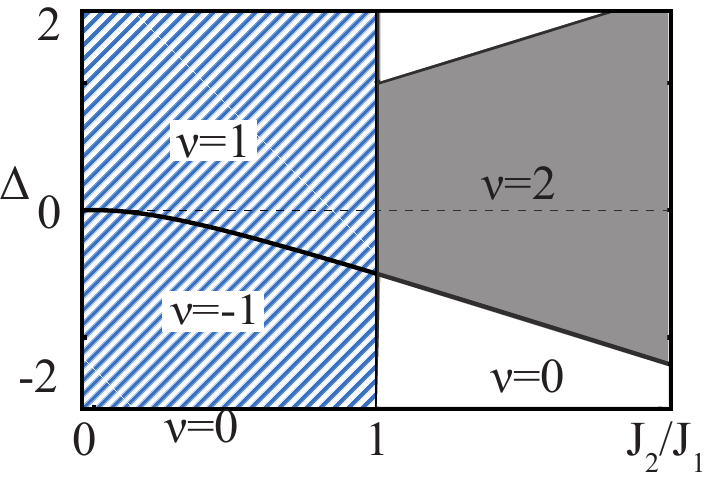}
\end{center}
\caption{Topological phase diagram in the chiral class for the case
$\left\vert \Delta_{1}\right\vert \neq J_{1}$ but still $\Delta_{2}=J_{2}$ and
$\mu=0$. We see that we can obtain regions with $\nu= 0$ (white),
$1$ (striped) and $2$ (grey). \label{fig:Phasediagram_D}}%
\end{figure}

For $\left\vert \Delta_{\alpha}\right\vert =J_{\alpha}$ and $\mu=0$, the gap
closing condition corresponds to $J_{1}=J_{2}$, and straightforward
calculations give $\nu_{\phi=0}=\frac{1}{2}\left[  3-\mathrm{sign}\left(
\frac{J_{1}-J_{2}}{J_{1}+J_{2}}\right)  \right]  $ and $\nu_{\phi=\pi}%
=\frac{1}{2}\left[1-  3\mathrm{sign}\left(  \frac{J_{1}-J_{2}}{J_{1}+J_{2}%
}\right)\right]  $. For $J_{1}=J_{2}$ we indeed have a topological phase
transition between two different states. For a finite chain, the winding
number manifests itself in the number of zero-energy modes localized near the
chain edges, see Figs. \ref{fig:Edgemodes_BDI}a and \ref{fig:Edgemodes_BDI}b. When
one approaches the topological phase transition point, say, from the side
$J_{2}>J_{1}$, two out of four zero-energy modes start to proliferate inside
the bulk of the chain and become gapped on the other side of the transition;
see Fig.~\ref{fig:Edgemodes_BDI}c.

The topological phase diagram in this chiral class for the case
$\left\vert \Delta_{1}\right\vert \neq J_{1}$, but still $\Delta_{2}=J_{2}$ and
$\mu=0$, is shown in Fig. \ref{fig:Phasediagram_D}. In this case $\vec{h}%
_{k}=(0,\Delta_{1}\sin k+J_{2}\sin2k,J_{1}\cos k+J_{2}\cos2k)$, where we
assume $\phi_{2}=0$, and, hence, real $\Delta_{1}$. The boundaries between
different topological phases results from the condition $\left\vert \vec
{h}_{k}\right\vert =0$.

For the \textrm{D} class [when $\phi\neq0,\pi(\mathrm{mod}2\pi)$], there
are only two different classes of the topological states characterized by the
Chern parity number $P_{C}\in\mathbb{Z}_{2}$. To define this topological invariant [4], one
has to extend the Hamiltonian $\mathcal{H}_{k}$ and thus $\vec{h}_{k}$ from a
one-dimensional Brillouin zone $k\in\lbrack-\pi,\pi]\sim S^{1}$ (topologically
equivalent to the circle $S^{1}$) to a two-dimensional ($2D$) Brillouin zone
$k,t\in\lbrack-\pi,\pi]\sim S^{1}\times S^{1}=T^{2}$ (topologically equivalent
to a two-dimensional torus $T^{2}$), $\mathcal{H}_{k}\rightarrow
\mathcal{H}_{k,t}=$ $\vec{h}_{k,t}\cdot\vec{\sigma}$, such that the resulting
Hamiltonian is gapped and belongs to the \textrm{D} class in $2D$: $\sigma
_{x}\mathcal{H}_{-k,-t}^{\ast}\sigma_{x}=-\mathcal{H}_{k,t}$ \footnote{Note that one can alternatively use the Pfaffian invariant defined in Ref.~\cite{Kitaevchain} to obatin the topological invariant.}. The unit vector
$\vec{n}_{k,t}=\vec{h}_{k,t}/|\vec{h}_{k,t}|$ maps then the $2D$ Brillouin
zone $T^{2}$ into a $2D$ sphere $S^{2}$. Different topological classes of such
mappings are distinguished by the integer-valued Chern number%
\[
C=\int_{-\pi}^{\pi}\frac{dkdt}{4\pi}\vec{n}_{k,t}\cdot(\partial_{k}\vec
{n}_{k,t}\times\partial_{t}\vec{n}_{k,t})\in\mathbb{Z}.
\]
It turns out [4], that the parity $P_{C}$ of $C$ does
not depend on the chosen extension of the Hamiltonian, and, therefore,
provides the topological invariant for $1D$ Hamiltonians in the \textrm{D}
class. (We set $P_{C}=1$ for odd $C$ and $P_{C}=0$ for even $C$. Topologically
nontrivial states correspond to $P_{C}=1$.)

\begin{figure}[t]
\begin{center}
\includegraphics[width=0.95\columnwidth]{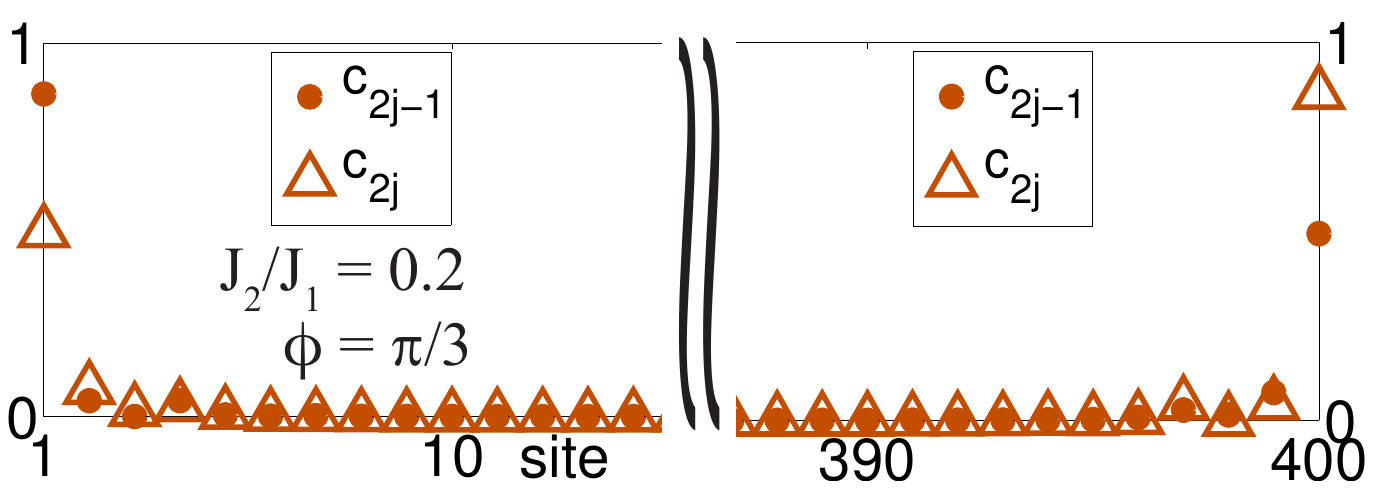}
\end{center}
\caption{Spatial distribution of the zero modes for $\phi=\pi/3$ and
$J_2/J_1=0.2$. There are two zero modes (left and right) which are linear combinations of only even or only odd Majorana operators living on a site $j$.}%
\label{fig:edgemodes_general}%
\end{figure}

The extension can be constructed using a hermitian matrix $\bar{\mathfrak{h}%
}(k,t)=\sum_{i}\mathfrak{h}_{i}(k,t)\sigma_{i}$, which has a gapped spectrum
and depends on two parameters $k\in\lbrack-\pi,\pi]$ and $t\in\lbrack0,\pi]$
such that $\bar{\mathfrak{h}}(k,0)=\mathcal{H}_{k}$ and $\bar{\mathfrak{h}%
}(k,\pi)=|\vec{h}_{k}|\sigma_{z}$. Note that $\bar{\mathfrak{h}}(k,t)$
interpolates between the initial Hamiltonian $\mathcal{H}_{k}$ and the
"trivial" one $|\vec{h}_{k}|\sigma_{z}$, both from the \textrm{D} class (for
$t\neq0$ or $\pi$, $\bar{\mathfrak{h}}(k,t)$ does not necessarily belong to
the \textrm{D} class in $1D$). Such a matrix always exists because it belongs
to the \textrm{A }class of general hermitian matrices (with no symmetries)
with a gapped spectrum that is topologically trivial. Then the matrix%
\begin{equation}
\mathcal{H}_{k,t}=\cases{\bar{\mathfrak{h}}(k,\pi+t), & for\; $t\in[-\pi,0]$\\
-\sigma_{1}\bar{\mathfrak{h}}^{T}(-k,\pi-t)\sigma_{1} & for $t\in[0,\pi]$
\\}
\end{equation}
is in the \textrm{D} class and provides the desired extension. The
interpolation $\bar{\mathfrak{h}}(k,t)$ can be obtained, for example, in the
following way: $\mathfrak{h}_{i}(k,t)$ for $t\in\lbrack0,\pi/2]$ corresponds
to a rotation of $\vec{h}_{k}=\mathfrak{h}_{i}(k,0)$ to the $y$-axes, such
that $\mathfrak{h}_{i}(k,\pi/2)$ is parallel to the $y$-axes for all $k$, and
then $\bar{\mathfrak{h}}(k,t)$ for\ $t\in\lbrack\pi/2,\pi]$ describes the
rotation of $\mathfrak{h}_{i}(k,\pi/2)$ to the $z$-axis.

Following this strategy for the case $\left\vert \Delta_{\alpha}\right\vert
=J_{\alpha}$ with $\mu=0$ we find $C=0$ for $J_{2}>J_{1}$ and $C=\mathrm{sign}%
(\sin\phi)$ for $J_{1}>J_{2}$, where $\phi=\phi_{1}-\phi_{2}$. The point
$J_{1}=J_{2}$ corresponds to the topological phase transition, at which the
gap vanishes, see Fig.~\ref{fig:Phase_diagram_allphi}c.

\subsection{AMO realization of the Zig-zag Chain}\label{Sec_pairingterm}
\begin{figure}[t]
\begin{center}
\includegraphics[width=0.95\columnwidth]{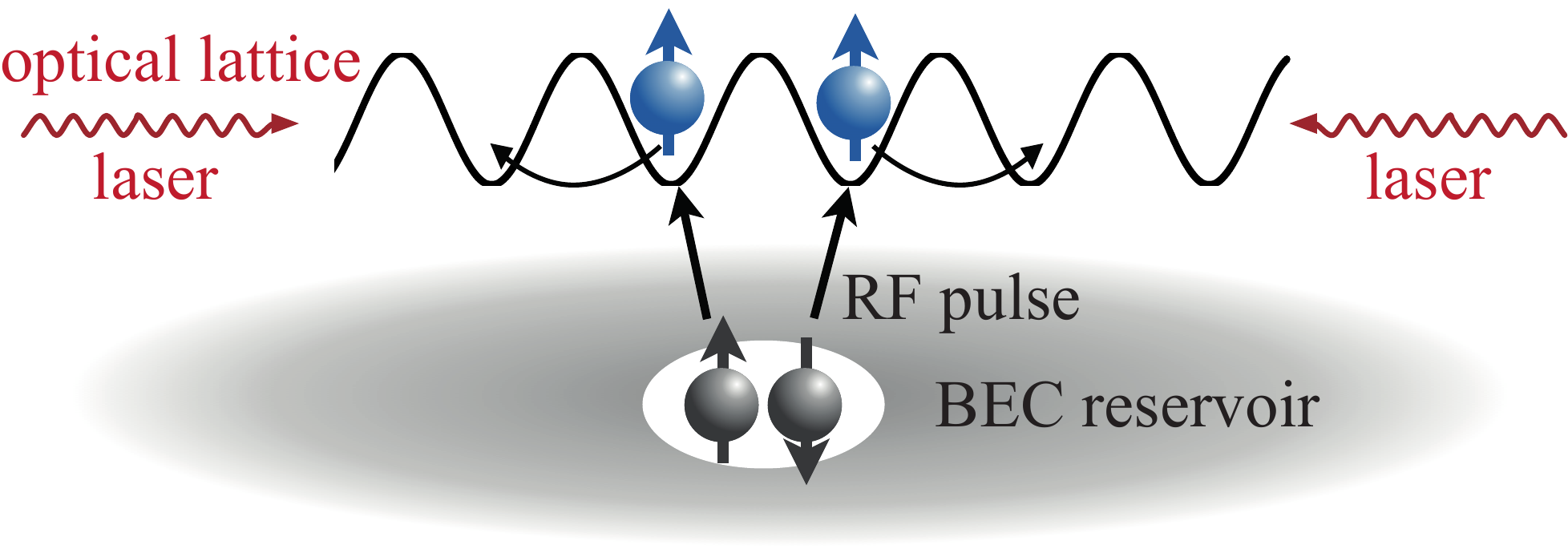}
\end{center}
\caption{Creation of the pairing term $\Delta \ad_j \ad_{j+1} + h.c.$ according to Ref~\cite{Probing_Liang}.  A one-dimensional system of trapped fermions in two spin states is coupled to a molecular BEC via a RF pulse. An additional Raman laser with strong driving projects out one of the spin components, leading to the desired pairing term.}%
\label{fig:create_wire}%
\end{figure}

The zig-zag chain defined in Eq. (\ref{eq:H_zigzag}) allows for an optical lattice realization, as we briefly explain in this Subsection (see also Fig.~\ref{fig:create_wire}). While the hopping term $\ad_j a_{j+\alpha} + h.c.$ arises naturally in an optical lattice setup, the pairing term $\ad_j \ad_{j+\alpha} + h.c.$ can be engineered via the coupling of the system to a BEC reservoir of Feshbach molecules, as explained in~\cite{Probing_Liang}. The main idea is to couple the two internal spin states of the trapped $1D$ system of fermions to a Feshbach molecule via an RF pulse. If we denote by $(\ad_{p, \uparrow}, \ad_{p, \downarrow})$ the two internal states of the trapped atoms with momentum $p$, then the result of the RF pulse is an effective pairing term of the form $\Delta \ad_{p, \uparrow}\ad_{-p, \downarrow}+h.c$. Driving the lattice fermions additionally with a Raman laser creates an effective magnetic field, which for strong driving (i.e. large magnetic fields) projects out one of the spin components, such that we obtain the spinless pairing term of Eq. (\ref{eq:H_zigzag}) ~\cite{Probing_Liang}.

\section{Preparation of Majorana modes}\label{Sec_preparation}
\begin{figure}[t]
\begin{center}
\includegraphics[width=0.95\columnwidth]{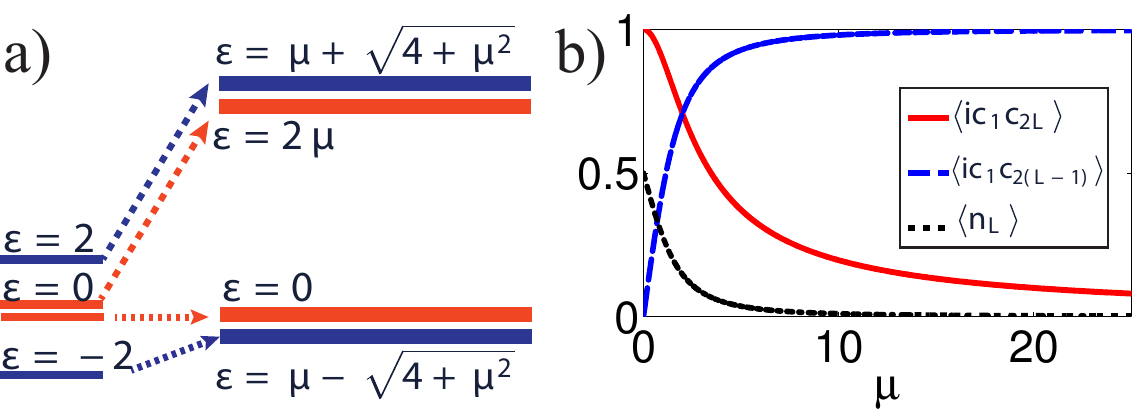}
\end{center}
\caption{a)Evolution of energy spectrum of the Hamiltonian $H(\mu)=H_{K}+\mu
(t)a_{L}^{\dagger}a_{L}|_{j=1,L}$ for an adiabatic ramp of $\mu$. b) Evolution of the occupation of the site $L$  and the Majorana correlation functions during the adiabatic ramp.}%
\label{fig:create_supp}%
\end{figure}
The main difficulty in detecting the Majorana modes is
that their signal comes only from the edges and, therefore, has poor
signal-to-noise ratio. To overcome this, an experiment has to be performed
several times thus requiring a reproducible preparation of a desired initial
quantum state of an open wire. In our setup, this can be achieved by starting
with a \textsl{closed} quantum chain in a fully paired \cite{Zwierlein_review,
Leggett} (and, hence, unique) state corresponding to the ground state of a
Hamiltonian (see Fig. 1b)). This state has an \textsl{even parity} (all
particles are paired) and only local correlations between the sites. Making
use of local addressing \cite{Bloch_singlespin, Greiner_singlesite} we can
"cut" the chain  by ramping up the chemical
potential on the site $L$,
\[
H(\mu)=H+\mu(t)a_{L}^{\dagger}a_{L}.
\]
This process affects only the four Majorana modes $c_{2L-2},c_{2L-1}%
,c_{2L},c_{1}$, and the corresponding instantaneous eigenvalues are $0$ and
$2\mu$ for the odd parity sector, and $\mu\pm\sqrt{4+\mu^{2}}$ for the even
one, such that there are two degenerate ground states of the Kitaev chain for
$\mu\rightarrow\infty$. Because a superposition of fermionic states with
different parity is forbidden by superselection rules, the corresponding
eigenstates never mix during the adiabatic ramp (cf.
Fig.~\ref{fig:create_supp}a), and, hence, the final state of the open chain
has even parity. In Fig. \ref{fig:create_supp}b we depict the evolution of the Majorana correlation functions between nearest- ($\langle ic_1c_{2L}\rangle$, solid) and next-to-nearest ($\langle ic_1c_{2L-1}\rangle$, dashed) neighbors, as well as the occupation of the site $L$ ($\langle n_L\rangle$, dotted) during the adiabatic ramp. 

For the "zig-zag" chain we can start with two decoupled closed chains (with
even and odd lattice sites) when only $\Delta_{2}$ and $J_{2}$ are nonzero but
$\Delta_{1}=J_{1}=0$. We then cut them as described above into two open chains
each with two edge Majorana modes and in the even fermionic parity state with
all fermions being paired. Now we can switch $J_{1}$ and $\Delta_{1}$ on
adiabatically to create a single quantum chain with even parity. Note that
during this process there is no single-particle transition between the two
subchains (with even and odd sites) because such a transfer would create
single-particle excitations in both subchains which cannot occur in an adiabatic process due to the conservation of energy. Thus, each of the subchains remains in the even parity state.

Further increase of NN amplitudes results in a topological phase transition with
closing the single-particle excitation gap allowing for a single-particle exchange
between the subchains. Then, only the (even) parity of the entire
chain is preserved. We would like to add here that parity violating processes, such as three-body losses, are sufficiently weak in fermionic systems (0.1--1s) in order not to be an experimental obstacle for the realization of Majorana physics

Note that a similar idea can be used to create the initial Majorana
state in the dissipative setting by adiabatically emptying one site ~\cite{Diehl_Natphys}.
Note that the cutting procedure avoids the need to establish the edge-edge
correlations via a sequential process, which takes polynomial time due to the
Lieb-Robinson bound \cite{Verstraete06,Eisert06}. 
\begin{figure}[ptb]
\includegraphics[width=0.95\columnwidth]{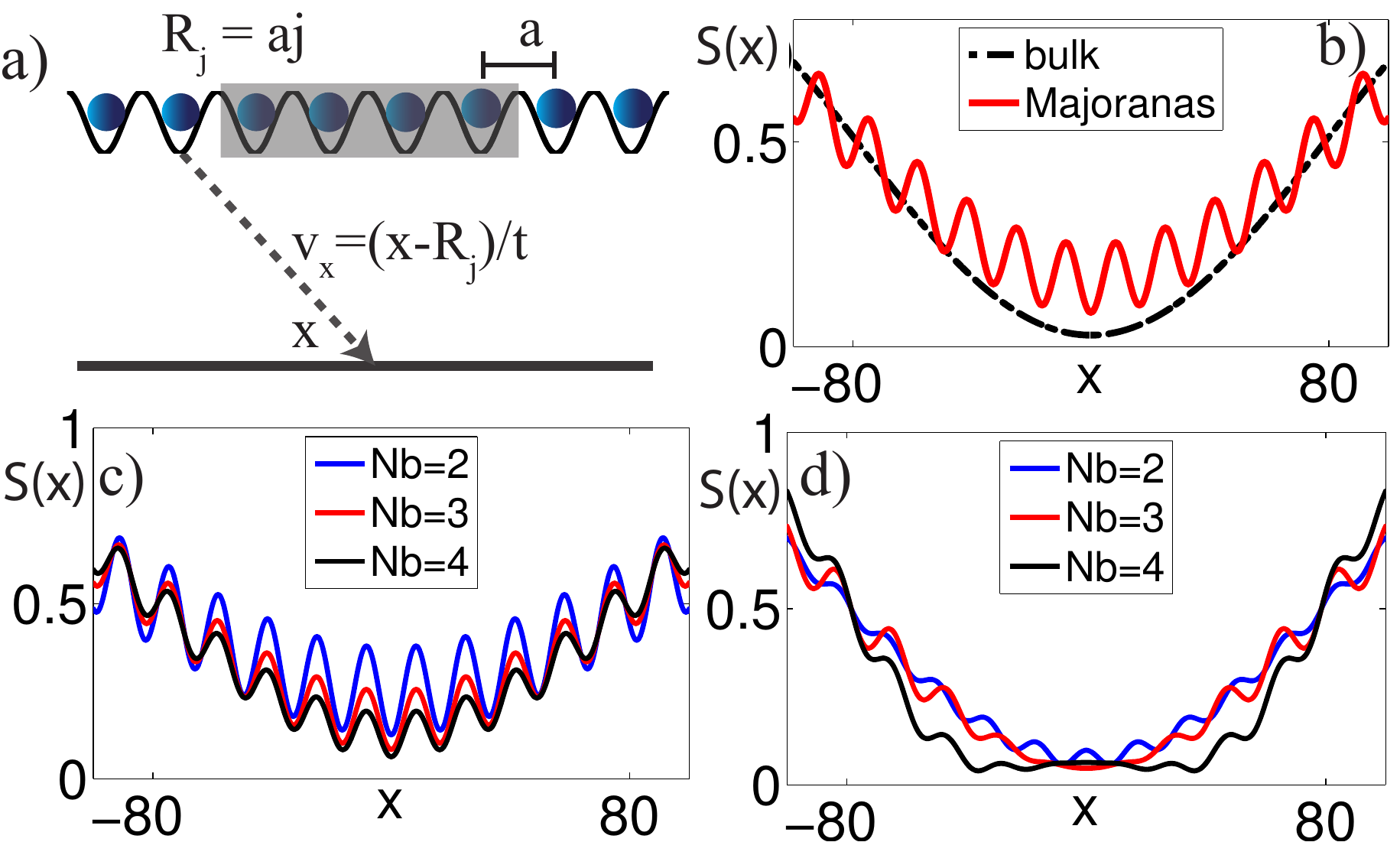} \caption{a) Principle of TOF. The contrast of the signal can be enhanced by
excluding the bulk (grey shade). b) TOF for the ideal Kitaev chain of length
$L_{M} = 20$: The presence of the Majorana modes (solid) leads to oscillations
of the signal $S(x)$ (see main text) which are absent in the bulk (dashed).
Having their origin in nonlocal fermionic correlations of modes located at the
edges, the oscillation frequency is independent of the size of the shade (c)
which allows to rule out a finite size effect encountered, e.g. in a system of
freely hopping particles (d) (see main text).}%
\label{fig:TOFmainone}
\end{figure}

\section{Time-of-Flight (TOF) imaging}\label{Sec_TOF}
Having prepared the atomic Majorana fermions following the protocol presented in Sec.~\ref{Sec_preparation}, we present now two possibilities to detect them in AMO setups. The first approach that is discussed in this Section is based on Time-of-Flight imaging allows to detect the existence and number of Majorana fermions in an optical lattice setup. Further, we present a complementary approach based on a spectroscopic setup in Sec.~ \ref{Sec_spectroscopy}.

Non-local correlations can be detected in TOF imaging, as illustrated in Fig.~\ref{fig:TOFmainone}a. We consider atoms
of mass $m$ released at time $t=0$ from lattice sites $R_{j}=ja$ with lattice
spacing $a$. At time $t\gg ma^{2}/\hbar$ (far field approximation) the atomic
density distribution at the detector is given by $\langle n(x)\rangle\sim
\sum_{j,j^{\prime}}e^{2ix(R_{j}-R_{j^{\prime}})m/(\hbar t)}\langle
a_{j}^{\dagger}a_{j^{\prime}}\rangle$ revealing the initial correlations of
the atoms in the lattice. Therefore, the long-range Majorana correlations
$R_{j}-R_{j^{\prime}}\approx L_{M}$ (distance between the edge modes) will
result in rapid oscillations of $\langle n(x)\rangle$ as compared to slow
oscillations originating from short-range bulk correlations ($R_{j}%
-R_{j^{\prime}}\approx a$). The contrast of the Majorana signal can be
enhanced by local addressing: we shade all but $N_{b}$ sites adjacent to each
of the two edges ($N_{b}=2$ in Fig.~\ref{fig:TOFmainone}a), thus measuring
$\langle n(x)\rangle= 2N_{b} S(x)$, where we have normalized by the number of
sites contributing to the signal, $2N_{b}$.

 \begin{figure}[tb]
\includegraphics[width=0.95\columnwidth]{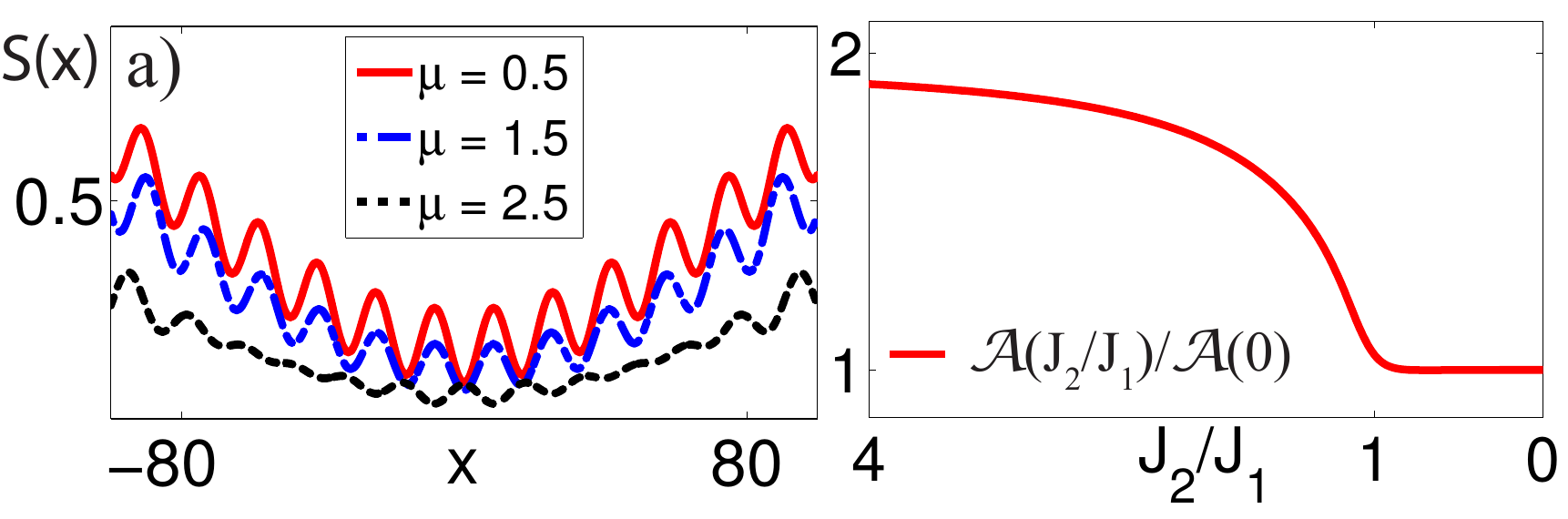} \caption{TOF as a method for detecting topological phase transitions: a) The
TOF signal for a single wire vanishes if we approach the transition point at
$\mu=2$. d) The amplitude of $S(x)$ allows to determine the number of Majorana
modes in the BDI class of our model (see main text).\label{fig:TOFmaintwo}}%
\end{figure}

The calculation of the density distribution at the detector in a TOF
experiment  requires the knowledge of the first moments
$\langle a_{j}^{\dagger}a_{j^{\prime}}\rangle$ of the state in the lattice
before the trapping potential is switched off. For $J_{\alpha}=\Delta_{\alpha}$, $J_{2}=\mu=0$ (ideal Kitaev chain), the
non-vanishing first moments in the Majorana language are $\langle c_{2j}%
c_{2k}\rangle_{\pm}=\langle c_{2j+1}c_{2k+1}\rangle_{\pm}=\delta_{kj}$,
$\langle c_{2j}c_{2k+1}\rangle_{\pm}=-\langle c_{2j+1}c_{2k}\rangle_{\pm
}=-i\delta_{kj}$, and the edge-edge correlations are given by $\langle c_{2L}c_{1}\rangle_{\pm}=im_{\pm}(d)$ [$m(d)=0$ if we consider the bulk only] with $d$
being the distance between the two Majorana edge modes. When only $N_{b}$
sites on the right and on the left sides of the chain contribute to the signal
(other sites are covered with a shade), the atom density distribution at the
detector reads
\[
\langle n(x)\rangle_{+}\sim\mathcal{N}_{bulk}(x)+m(d)\mathcal{N}_{edge}(x),
\]
where $\mathcal{N}_{bulk}(x)=2N_{b}\sin^{2}(\frac{mxa}{\hbar t})+\cos
(\frac{mxa}{\hbar t})$ stems from the atoms in the bulk only and the two edge
sites separated by $L_{M}=ad$ give rise to $\mathcal{N}_{edge}(x)=\cos
(\frac{mxa}{\hbar t}d)/2$ multiplied by the Majorana correlation
$m(d)=-i\left\langle \gamma_{L}\gamma_{R}\right\rangle $. For the general case, we obtain the correlation matrix $\langle \ad_j a_{j'}\rangle$ via a numerical diagonalization of the Hamiltonian $H$. 

We see that the second rapidly oscillating term reflects the presence of the
long-range Majorana correlations and allows to determine the occupation of the
Majorana subspace. If the initial state of the chain is prepared as described
above, then $m(d)=-i\left\langle G_{+}\right\vert \gamma_{L}\gamma
_{R}\left\vert G_{+}\right\rangle =1$, and the amplitude of this terms reaches
its maximum. The results of the calculations for the ideal chain are shown in
Fig.~\ref{fig:TOFmainone}b: The presence of the Majorana modes leads to
oscillations that are absent in the bulk. Having their roots in long-range
fermionic correlations, we can distinguish these oscillations from those
resulting from a finite size sample by changing the size of the shade: A
signal indicating the presence of Majorana fermions exhibits oscillations with
the same frequency but reduced amplitude when we include more sites of the
bulk (see Fig.~\ref{fig:TOFmainone}c). On the contrary, oscillations induced
by a finite size effect have in general a frequency that is sensitive to the
shape of the shade (see Fig.~\ref{fig:TOFmainone}d where we consider a system
of freely hopping fermions).

By ramping the chemical potential, we can detect the location of the phase
transition, since the oscillations disappear when crossing the transition
point at $\mu=2$ (see Fig.~\ref{fig:TOFmaintwo}a). Further, from the amplitude
$\mathcal{A}$ of the signal $S(x)$ we can deduce the number of Majorana modes,
as depicted in Fig.~\ref{fig:TOFmaintwo}b: The amplitude for four zero modes
($J_{2}>J_{1}$) drops by a factor of two upon reaching the transition point
$J_{2}=J_{1}$.

\begin{figure}[tb]
\includegraphics[width=0.95\columnwidth]{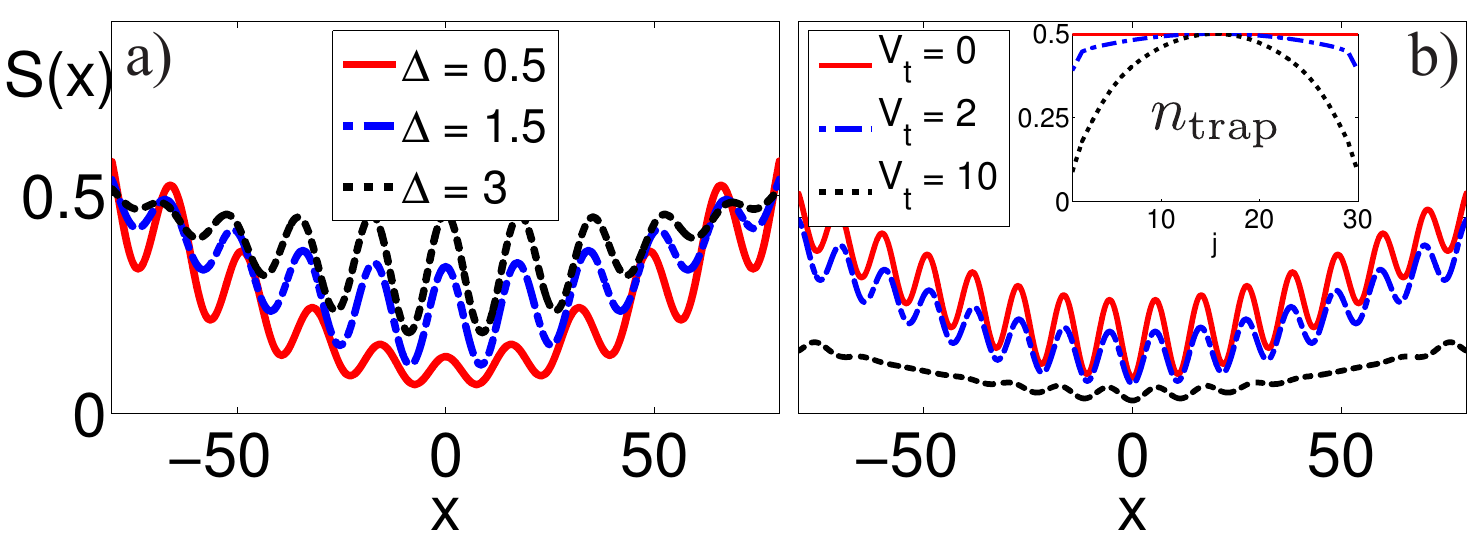} \caption{a) TOF for a non-ideal Kitaev chain where $J \neq \Delta$. b) Influence of an external harmonic trapping potential $H_{\mathrm{trap}} = -V_t\sum_x ((L-1)/2 - x)^2\ad_x a_x$ on the TOF signal for a system of $L=30$ sites and $\Delta = 1.2 J$. In the inset we show the density for different values of $V_t$ in units of $J$. As long as the density of the system is not too inhomogeneous only the amplitude of the signal is affected by the external confinement.  (see main text).\label{fig:TOF_imperfections}}%
\end{figure}

Finally, we also provide some results for experimentally realistic situations like a non-ideal chain and the influence of an external confinement of the atoms. In Fig. \ref{fig:TOF_imperfections} a) we show TOF signals for the non-ideal case where $J \neq \Delta$, proving that the oscillations will still be present in such a scenario. Further, we show the effect of an external harmonic trapping potential modeled by $H_{\mathrm{trap}} = -V_t\sum_x ((L-1)/2 - x)^2\ad_x a_x$ for a system of $L=30$ sites and $\Delta = 1.2$ in Fig. \ref{fig:TOF_imperfections} b). In the inset we present the local density distribution of the atoms in the trap. We find that in the case of not too large inhomogeneities the TOF experiment can be used for the detection of Majorana fermions.
  
\section{Spectroscopy}\label{Sec_spectroscopy}
\begin{figure}[bt]
\includegraphics[width=0.9\columnwidth]{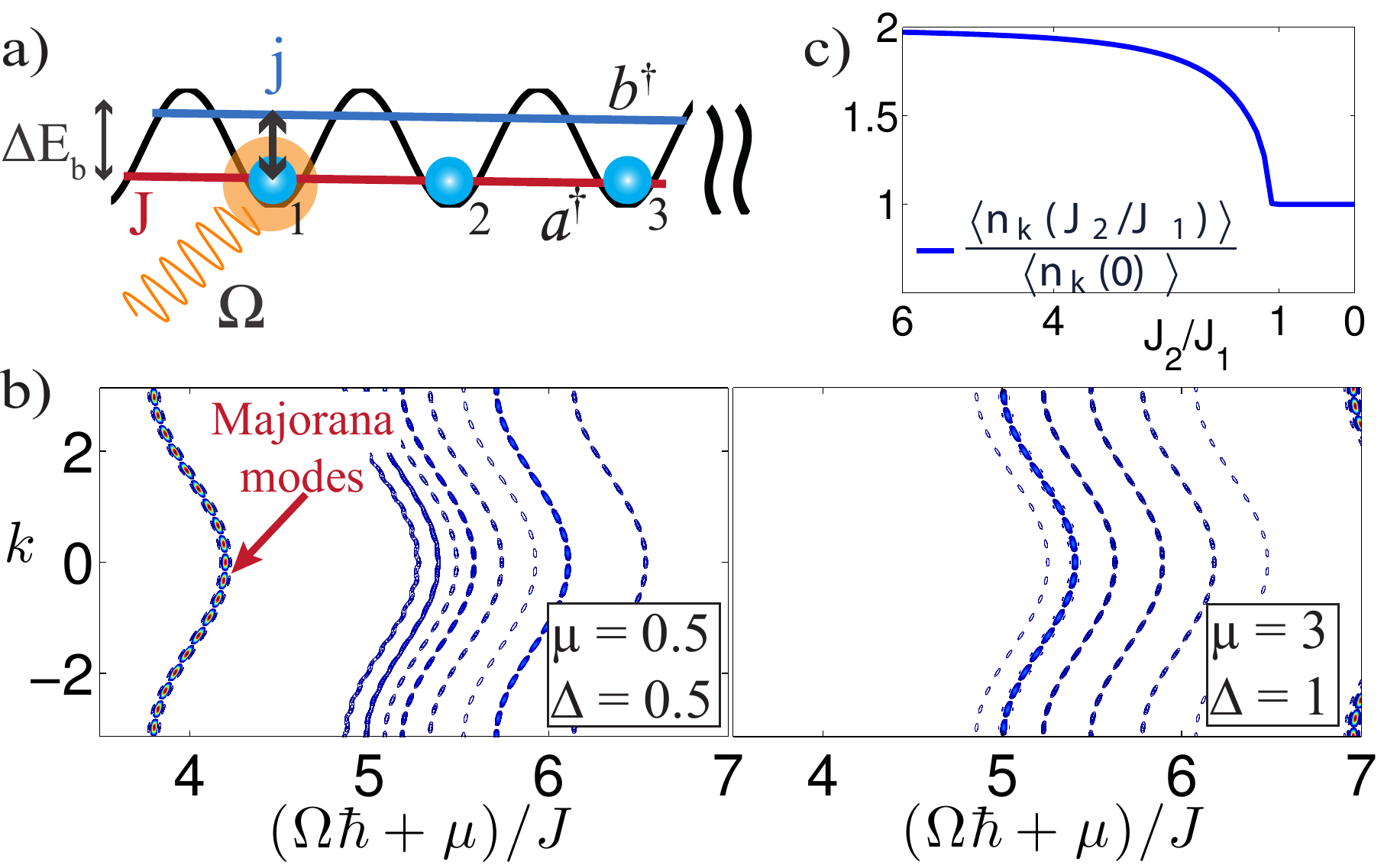} \caption{a) Spectroscopic detection of Majorana modes in a two-band lattice
(see text). b) The momentum distribution of atoms in the upper band ($\Delta
E_{b}=4J,j=0.1J$) reveals the presence (left) or absence (right) of Majorana
modes (see text). c) From the strength of the signal $\langle n_{k}\rangle$ at
resonance (here: $k=\pi/2$) we can deduce the number of Majorana modes.\label{fig:Spectroscopy}}
\end{figure}

While TOF imaging allows to detect the existence and number of Majorana fermions in an AMO setup, 
the combination of a spectroscopic setup with TOF allows
to measure not only the energy of the Majorana states but also their wave
functions. (The use of spectroscopic RF absorption spectra to detect Majorana
modes in $2D$ $p$-wave superconductors was considered in Ref.
\cite{Probing_Cooper_Auerbach}.) Consider a $1D$ lattice with two bands [modes
$a_{j}$ ($b_{j}$) in the upper (lower) band] separated by an energy difference
$\Delta E_{b}$. In the lower band we realize the Hamiltonian $H$, while the
fermions can hop freely with the hopping amplitude $j$ in the upper band. The
full Hamiltonian reads $H_{0}=H+H_{up}$, where $H_{up}=j\sum_{l=1}^{L}%
(b_{l}^{\dagger}b_{l+1}+h.c.)+\Delta E_{b}\sum_{l=1}^{L}b_{l}^{\dagger}b_{l}$.
We then couple upper and lower bands on the \textsl{first} and the
\textsl{last} sites of an open chain via a time-dependent perturbation
$V(t)=V_{0}(b_{1}^{\dagger}a_{1}+b_{L}^{\dagger}a_{L})e^{-i\Omega t}+h.c.$
(see Fig.~\ref{fig:Spectroscopy}a) and measure the momentum distribution in
the upper band as a function of the frequency $\Omega$, e.g., via TOF imaging.

Let us assume first that $J_2=\Delta_2 = 0, J_1 = \Delta_1$ (ideal Kitaev chain) in the lower band. Then, the Hamiltonian $H$ is diagonal in the Bogoliubov quasiparticle basis
$\tilde{a}_{l}=(a_{l}^{\dagger}-a_{l}+a_{l+1}^{\dagger}-a_{l+1})/2=(c_{2l+1}%
+ic_{2l})/2$, $H=2J_{1}\sum_{l=1}^{L-1}\tilde{a}_{l}^{\dagger}\tilde{a}%
_{l}$, and its ground state is two-fold degenerate, with the ground states
$|G_{+}\rangle$ and $|G_{-}\rangle$ having different fermionic parity.

We can now rewrite the external perturbation in terms of quasiparticle
operators $\tilde{a}_{l}$ and Majorana operators $\gamma_{L}$, $\gamma_{R}$ using
$a_{1}=(\gamma_{L}+\tilde{a}_{1}^{\dagger}-\tilde{a}_{1})/2$, and $a_{L}=(\tilde
{a}_{L-1}^{\dagger}+\tilde{a}_{L-1}-i\gamma_{R})/2$. Choosing the even parity
ground state $|G_{+}\rangle$ as an initial state and using $\gamma_{L}|G
_{\pm}\rangle=|G_{\mp}\rangle$ and $\gamma_{R}|G_{\pm}\rangle=\pm i|G
_{\mp}\rangle$, a straightforward application of the standard time-dependent
perturbation theory results in the following the upper band momentum
distribution:
\begin{eqnarray}
&  \langle n_{k}(t,\Omega)\rangle=V_{0}^{2}L^{-1}(n_{0}+n_{1}),\\
&  n_{0}=A_{0}f_{t}(\varepsilon_{k}-\Omega\hbar),\;\;\;\;\;n_{1}%
=f_{t}(\varepsilon_{k}+2J-\Omega\hbar),\nonumber
\end{eqnarray}
where we have introduced the notation $A_{0}=[m_{+}(d)-m_{-}(d)]\sin^{2}%
\frac{k}{2}(L-1)+m_{-}(d)$ and $f_{t}(x)=4\sin^{2}(xt/2\hbar)/x^{2}$.
Further, $\varepsilon_{k}=\Delta E_{b}+2j\cos k$ is the dispersion in the
upper band and $m_{\pm}(d)$ denote the initial occupation of the even/odd
parity subspace in the lower band. The contribution $n_{0}$ corresponds to a
parity flip in the lower band (no energy cost) and the creation of an
excitation (particle) in the upper band (energy $\varepsilon_{k}$). Thus, the
absorption peak at $\hbar\Omega=\varepsilon_{k}$ in $n_{0}$ indicates the
ground state degeneracy. The edge-edge Majorana correlation length is encoded
in the oscillation period. The term $n_{1}$ results from the creation of
excitations in both upper (energy $\varepsilon_{k}$) and lower (energy $2J$)
band. The corresponding absorption peak is located at $\hbar\Omega
=2J+\varepsilon_{k}$, providing the direct measurement of the pairing energy
gap ($2J$ here) from the distance between the two peaks. Thus, the
topologically non-trivial phase leads to a clear signal at $\Omega\hbar=2J$
(for $\mu\neq0$ one has to replace $\hbar\Omega\rightarrow\hbar\Omega-\mu$,
cf. Fig.~\ref{fig:Spectroscopy}b). Note that this setup allows to detect the
presence of the zero energy Majorana subspace independently of its purity.
Further, the number of Majorana modes is encoded in the strength of the
resonance signal for a fixed momentum $k$. In Fig.~\ref{fig:Spectroscopy}c we
present the results for the Hamiltonian $H$ with $J_{\alpha}=|\Delta_{\alpha
}|$, $\phi=0$, where the perturbation is on sites $1,2,L-1$ and $L$. Starting
from a state with four Majorana modes, we reduce the ratio $J_{2}/J_{1}$. At
the transition point, the amplitude of the resonance signal $\langle
n_{k=\pi/2}\rangle$ drops to half its magnitude indicating the disappearance
of two of the Majorana modes. If we go away from the ideal case, the quasiparticle operators and the corresponding matrix elements can be determined numerically.

\begin{figure}[ptb]
\includegraphics[width = 0.9\columnwidth]{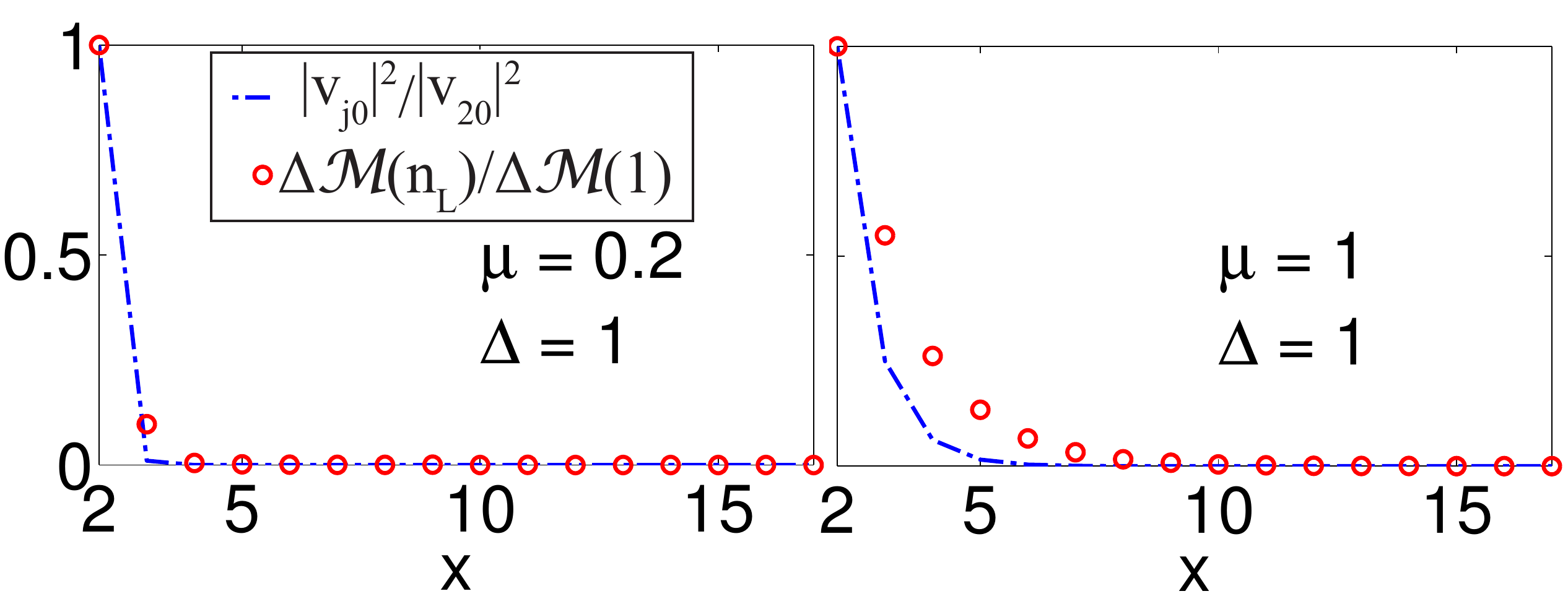}
\caption{Measurement of the localization length of the Majorana
mode via spectroscopy (see main text). }
\label{fig.localization}
\end{figure}
With a slight modification of the above setup one can also probe the
\textsl{localized character} of the Majorana zero modes. Namely, we apply the
perturbation to the first $n_{L}$ sites: $V^{(n_{L})}(t)=V_{0}\sum
_{j=1}^{n_{L}}a_{j}^{\dagger}b_{j}e^{i\Omega t}+h.c.$. The corresponding
momentum distribution in the upper band is now
\begin{equation}
\langle n_{k}({\Omega},t)\rangle=\frac{V_{0}^{2}}{L}\sum_{\nu=0}^{L-1}%
f_{t}(E_{\nu}+\epsilon_{k}-\Omega\hbar)\left\vert \sum_{j=1}^{n_{L}}%
e^{ikj}v_{j\nu}\right\vert ^{2},
\end{equation}
where $v_{jl}$ are the coefficients of the Bogoliubov transformation to the
quasiparticle operators $\tilde{a}_{\nu}$ ($\nu$ labels the quasiparticle
modes with energies $E_{\nu}$) in the lower band, $a_{i}=\sum_{j}(u_{i\nu
}\tilde{a}_{\nu}-v_{i\nu}^{\ast}\tilde{a}_{\nu}^{\dagger})$, which
diagonalizes $H$. Let us now consider $k=0$ and the external frequency being
in resonance with the lowest energy ($\nu=0$ Majorana) level, $\Omega
=\Omega_{r}=\Delta E_{b}+\epsilon_{k=0}$. In this case, only the Majorana mode
is excited in the lower band, and one has $\langle n_{k=0}(\Omega
_{r},t)\rangle\sim\left\vert \sum_{j=1}^{n_{L}}v_{j0}\right\vert ^{2}%
\equiv\mathcal{M}(n_{L})$. The \textsl{localized character} of the Majorana
mode can now be established by looking at the dependence of $\langle
n_{k=0}(\Omega_{r},t)\rangle$ on the number of sites $n_{L}$ affected by the
perturbation. To be more specific, the absolute change of $\mathcal{M}(n_{L})$
when $n_{L}$ is increased by one, $\Delta\mathcal{M}(n_{L})=\left\vert
\mathcal{M}(n_{L}+1)-\mathcal{M}(n_{L})\right\vert $, is expected to be
$\Delta\mathcal{M}(n_{L}>1)=0$ for the ideal Kitaev chain and $\Delta
\mathcal{M}(n_{L})\sim\exp(-n_{L}/l_{\mathrm{loc}})$ for a generic one. This
is reflected in a comparison of the normalized quantity $\Delta\mathcal{M}%
(n_{L})/\Delta\mathcal{M}(1)$ (red circles) with the normalized wave function
of the Majorana mode, $\left\vert v_{j0}\right\vert ^{2}/\left\vert
v_{20}\right\vert ^{2}$ (dashed) as shown in Fig.~\ref{fig.localization}.

\section{Probing Majorana fermions in $2D$.} 
\begin{figure}[t]
\begin{center}
\includegraphics[width =0.9\columnwidth]{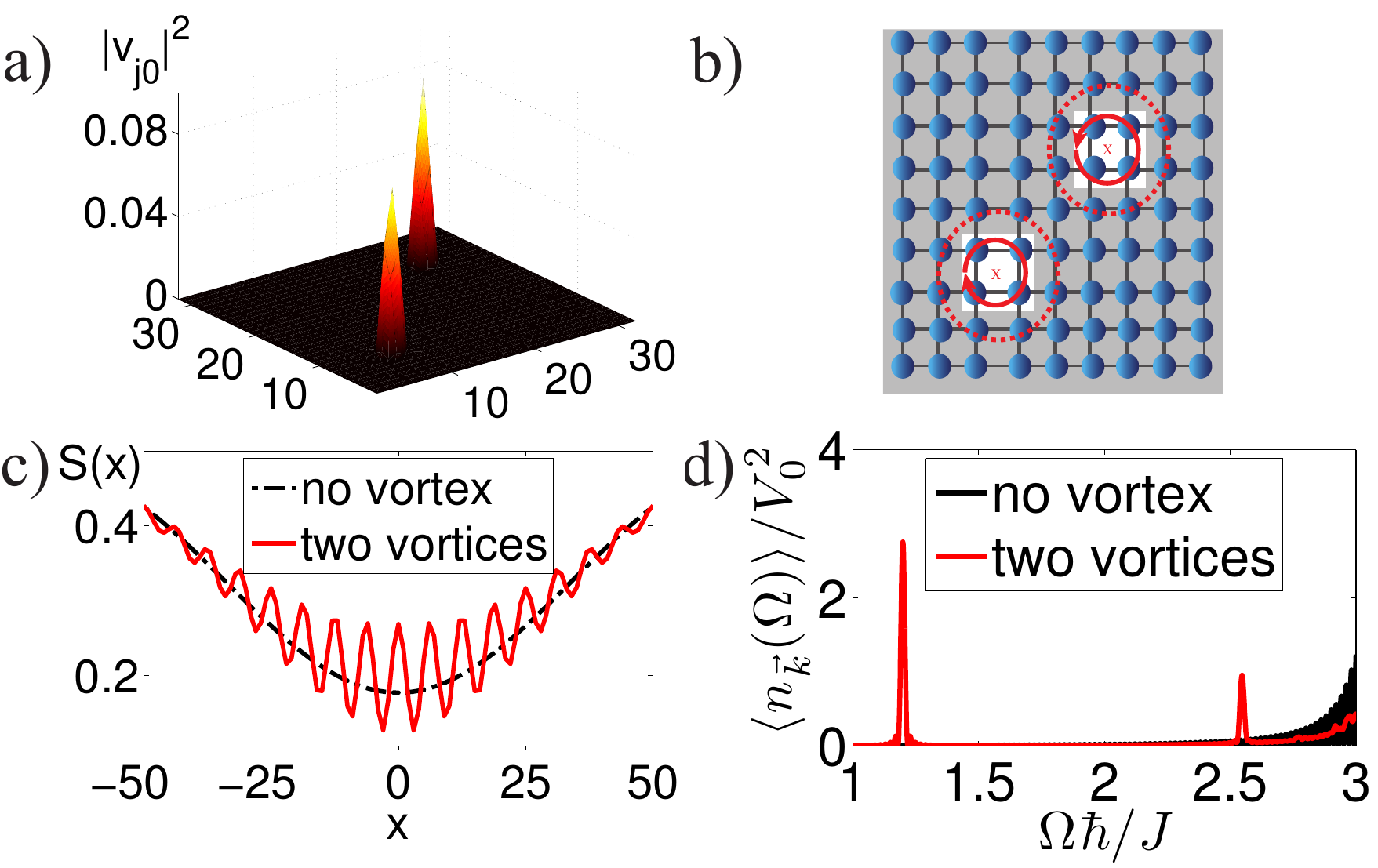}
\end{center}
\caption{a) Spatial distribution of the Majorana modes in the
vicinity of two vortices imprinted on a $p_{x}+ip_{y}$-superfluid for $\mu=2J$
on a $32\times32$ lattice. b) A shade (grey) is introduced to enhance the
signal contrast for the TOF (c) and spectroscopic (d) detections ($\Delta
E_{b}=J,j=0.1J$). }%
\label{fig:2d}%
\end{figure}

In the last Sections we have presented methods to prepare and detect Majorana fermions in a one-dimensional model system.
Note, however, that the proposed detection methods are universal
and can be applied to other systems. As an illustration, we consider a $2D$
$p_{x}+ip_{y}$ superfluid, as it was proposed in ~\cite{Ivanov, ReadGreen} with Hamiltonian
$H_{S}=H_{hopp}+H_{pair}+H_{\mu}$, with a NN hopping $H_{hopp}$, a pairing
term $H_{pair}=\sum_{\mathbf{x},\mathbf{y}}\Delta_{\mathbf{x},\mathbf{y}%
}a_{\mathbf{x}}^{\dagger}a_{\mathbf{y}}^{\dagger}+h.c.$ where $\Delta
_{\mathbf{x},\mathbf{y}}=\Delta_{x}(\delta_{\mathbf{y},\mathbf{x}%
+\mathbf{e}_{1}}+i\delta_{\mathbf{y},\mathbf{x}+\mathbf{e}_{2}})$, and
$H_{\mu}=-\mu\sum_{\mathbf{x}}a_{\mathbf{x}}^{\dagger}a_{\mathbf{x}}$. The
fermionic operators $a_{\mathbf{x}}$ are defined on a $2D$ (square) lattice
and the vectors $\mathbf{e}_{1,2}$ are two independent elementary lattice
translations. We introduce two vortices (each with vorticity $1$) via a
position dependent order parameter $\Delta_{\mathbf{x}}=|\Delta|e^{i\phi(x)}$,
which can be written onto the superfluid by phase imprinting in the Raman
process~\cite{Bloch_singlespin}.

For $\left\vert \mu\right\vert \ll4J$ we have a topological ground state
supporting two Majorana modes located in the vicinity of the two vortices (cf.
Fig.~\ref{fig:2d} a). The comparison of the TOF signal with and without
vortices (Fig.~\ref{fig:2d}c) shows that the presence of the Majorana modes
leads to an oscillatory behavior of the density $\langle n(x)\rangle$ at the
detector. Again, the contrast of the signal is enhanced by shading all but a
small region near the vortices (Fig.~\ref{fig:2d}b). To detect the Majorana
modes in the spectroscopic setup, we apply an external perturbation
$V(t)=V_{0}\sum_{\mathbf{x}\in\Lambda}a_{\mathbf{x}}b_{\mathbf{x}}^{\dagger
}e^{-i\Omega t}+h.c.$, where $\Lambda$ includes the lattice sites inside a
$w\times w$ small regions around the vortices (here: $w=3$). The resulting
momentum distribution in the upper band $\langle n_{k_{x},k_{y}}\rangle$ for
$k_{x,y}=\frac{\pi}{32}$ is depicted in Fig.~\ref{fig:2d}d. The presence of
the Majorana modes leads to a clear signal at $\Omega\hbar/J=\varepsilon
_{\frac{\pi}{32},\frac{\pi}{32}}/J\approx1.19$ that is absent in a
vortex-free setting.
\section{Summary}
In this work we have presented various techniques that allow for an unambiguous detection of atomic Majorana fermions and topological order with standard quantum optical tools, as time-of-flight imaging and spectroscopic techniques within our framework. To this end, we have introduced a one-dimensional model system that allows for an AMO realization and provides, due to its rich topological phase diagram, and ideal playground to test our ideas. Since some of our detection schemes require the preparation of a ground state with a definite parity, we have further provided a protocol that allows to achieve this goal. In addition, our detection techniques are universal in the sense that they can be used to detect atomic Majorana fermions in any dimension and geometry, as we have illustrated by the example of a $2D$ $p_x+ip_y$ superfluid.

\section{Acknowledgments}
We thank E. Rico for helpful discussions. We acknowledge support by the
Austrian Science Fund (FWF) through SFB FOQUS and the START grant Y 581-N16
(S. D.), the European Commission (AQUTE), the Institut fuer
Quanteninformation GmbH and the DARPA OLE program.

\end{document}